\documentclass{PoS}

\title{NLO QCD and EW corrections to processes involving off-shell top quarks}

\ShortTitle{NLO QCD and EW corrections to processes involving off-shell top quarks}

\author{Ansgar Denner, Jean-Nicolas Lang, \speaker{Mathieu Pellen}\\
        Universit\"at W\"urzburg, %
        Institut f\"ur Theoretische Physik und Astrophysik, %
        Emil-Hilb-Weg 22,
        97074 W\"urzburg, %
        Germany \\
        E-mail: \email{denner@physik.uni-wuerzburg.de}, \email{mpellen@physik.uni-wuerzburg.de}, \email{jlang@physik.uni-wuerzburg.de}}

\author{Sandro Uccirati \\
        Universit{$\grave{a}$} di Torino e INFN, 10125 Torino, Italy, \\
        E-mail: \email{uccirati@to.infn.it}}

\abstract{We review recent results on next-to-leading order (NLO) QCD and electroweak (EW) corrections for processes involving off-shell top quarks.
For the off-shell production of two top quarks that decay leptonically, the full NLO EW corrections have been computed.
For the very same process in association with a Higgs boson, EW corrections have been calculated and combined with existing NLO QCD corrections.
In both cases, selected differential distributions are shown.
In these proceedings, particular emphasis is put on the effect of the EW corrections as well as the off-shell contributions.}

\FullConference{13th International Symposium on Radiative Corrections (Applications of Quantum Field Theory to Phenomenology)\\
         25-29 September, 2017 \\
         St. Gilgen, Austria}

\begin{document}

\section{Introduction}

The study of the top quark is one of the major tasks of the experimental program at the Large Hadron Collider (LHC).
In particular, its interaction with the Higgs boson is actively being investigated.
The experimental precision allows to probe the tails of differential distributions.
In these tails where new physics is expected to appear, electroweak (EW) corrections as well as non-resonant contributions are particularly large, motivating their precise investigation.
In that respect, we have computed the NLO EW corrections to the hadronic processes ${\rm p p} \to {\rm e}^+ \nu_{\rm e} \mu^-  \bar{\nu}_\mu {\rm b} \bar{{\rm b}}$ \cite{Denner:2016jyo} and 
${\rm p p} \to {\rm e}^+ \nu_{\rm e} \mu^-  \bar{\nu}_\mu {\rm b} \bar{{\rm b}} {\rm H}$ \cite{Denner:2016wet}.
For the latter, the next-to-leading order (NLO) EW corrections have been combined with the QCD ones already computed in Ref.~\cite{Denner:2015yca}.
In this contribution, we only highlight some of the results presented in Refs.~\cite{Denner:2016jyo,Denner:2016wet}.
The input parameters and phase-space cuts are not reproduced here.
For more details and materials, we refer the reader to the original articles.

Both computations have been performed with the help of the same tools.
First, we have used the Monte Carlo {\sc MoCaNLO} which has been designed for the computation of NLO QCD corrections to the production of two top quarks decaying leptonically in association with a Higgs boson \cite{Denner:2015yca}.
It has proven to be particularly efficient for computing NLO corrections for high-multiplicity processes and is based on similar phase-space mappings as the ones of Ref.~\cite{Dittmaier:2002ap}.
The matrix elements have been exclusively obtained from the program {\sc Recola} \cite{Actis:2012qn,Actis:2016mpe} in association with the {\sc Collier} library \cite{Denner:2014gla,Denner:2016kdg}.
In particular, all the tree, one-loop, and colour-correlated amplitudes necessary for the computation have been obtained in a completely automatic way thanks to {\sc Recola}.
Finally, the infrared divergences have been handled with the help of the Catani--Seymour dipole formalism \cite{Catani:1996vz} and its extension to QED \cite{Dittmaier:1999mb}.

\section{Off-shell production of top-antitop pairs}

\begin{figure}

                \includegraphics[width=0.5\textwidth]{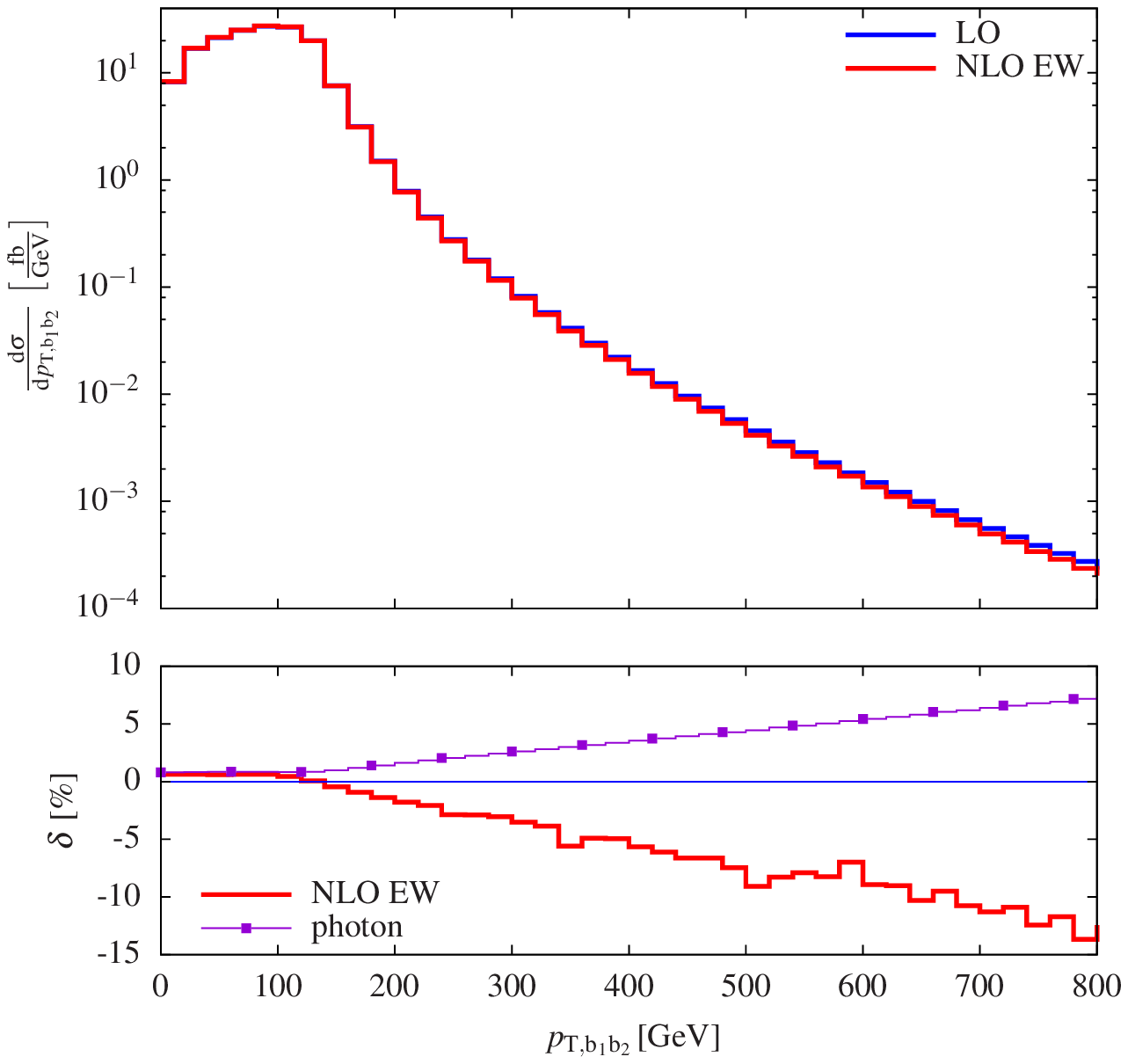}
                \includegraphics[width=0.5\textwidth]{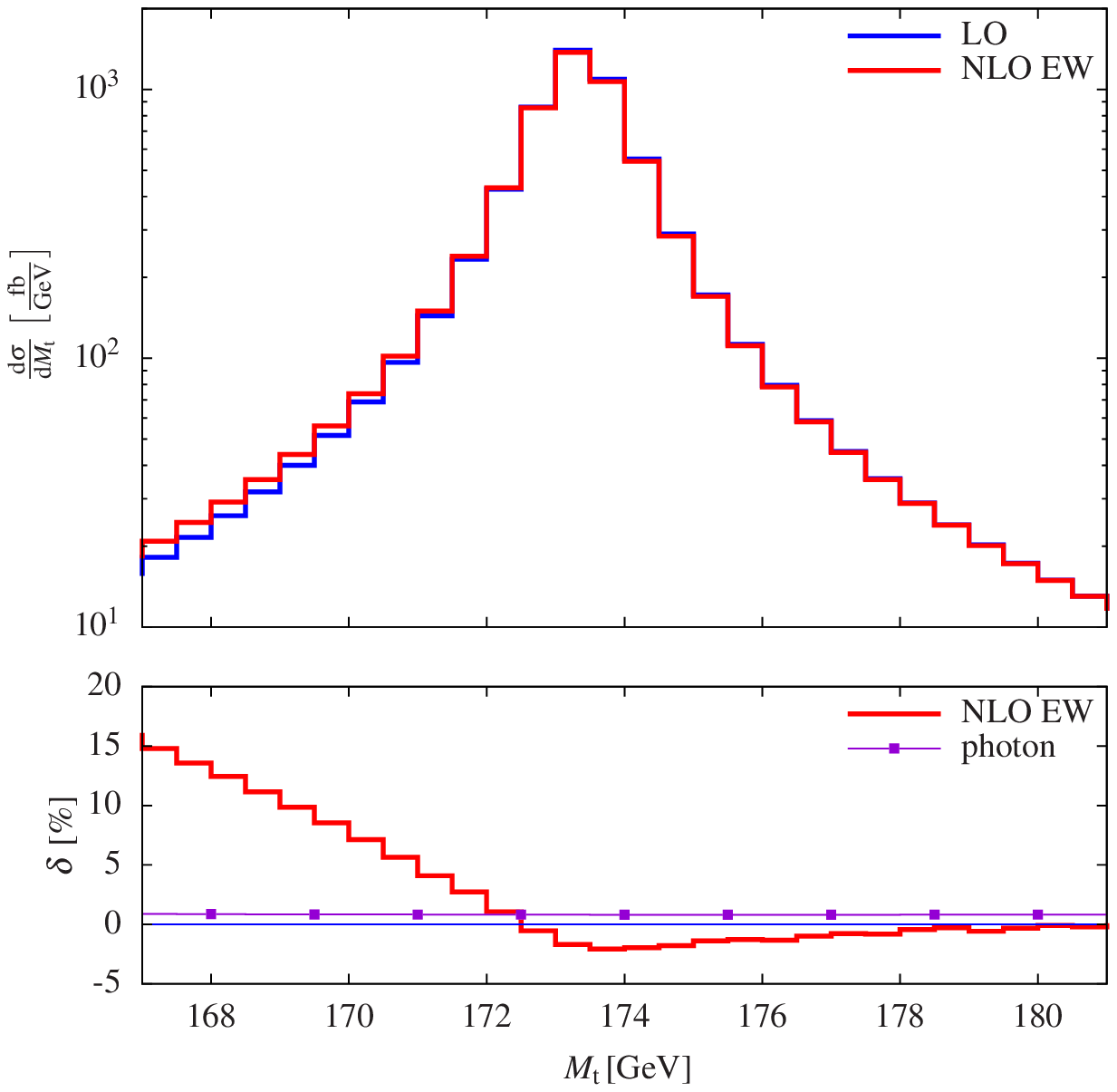}
	
        \caption{Differential distributions \cite{Denner:2016jyo} for ${\rm p p} \to {\rm e}^+ \nu_{\rm e} \mu^-  \bar{\nu}_\mu {\rm b} \bar{{\rm b}}$:
        transverse momentum of the two bottom quarks (left) and invariant mass of the top quark (right).
        The NLO EW corrections as well as the tree-level photon-induced contributions are shown.}
\label{fig:tt}
\end{figure}

The first process studied is ${\rm p p} \to {\rm e}^+ \nu_{\rm e} \mu^-  \bar{\nu}_\mu {\rm b} \bar{{\rm b}}$.
We consider leading order (LO) contributions of order $\alpha^2_{\rm s} \alpha^{4}$ and the corresponding NLO EW corrections of order $\alpha^2_{\rm s} \alpha^{5}$.
In Fig.~\ref{fig:tt}, the NLO EW corrections to the full process are displayed for the distribution in the transverse momentum of the two bottom jets as well as for the distribution in the invariant mass of the top quark.
In both distributions the NLO EW corrections are large but they do not have same origin.
In the transverse momentum distribution, they grow negatively large towards high transverse momentum owing to the appearance of Sudakov logarithms.
On the other hand, the radiative tail observed below the peak of the invariant-mass distribution is due to photon emission not reconstructed with the decay products of the top quark.
Such an effect has already been observed for NLO QCD corrections to top-quark production \cite{Denner:2010jp}.
In addition, the photon-induced contribution of the type ${\rm g}\gamma / \gamma{\rm g} \to {\rm e}^+ \nu_{\rm e} \mu^-  \bar{\nu}_\mu {\rm b} \bar{{\rm b}}$ at order $\alpha_{\rm s} \alpha^{6}$ at LO are also displayed in Fig.~\ref{fig:tt}.
Interestingly, one observes an increase of these contributions for high transverse momentum.
This has been confirmed in Ref.~\cite{Pagani:2016caq} for on-shell top quarks and is partly due to outdated photon distribution functions.
Using the present most precise parton distribution functions (PDF) LUXQED \cite{Manohar:2016nzj}, the photon contributions are smaller \cite{Denner:2016wet}.

\begin{figure}
                \includegraphics[width=0.5\textwidth]{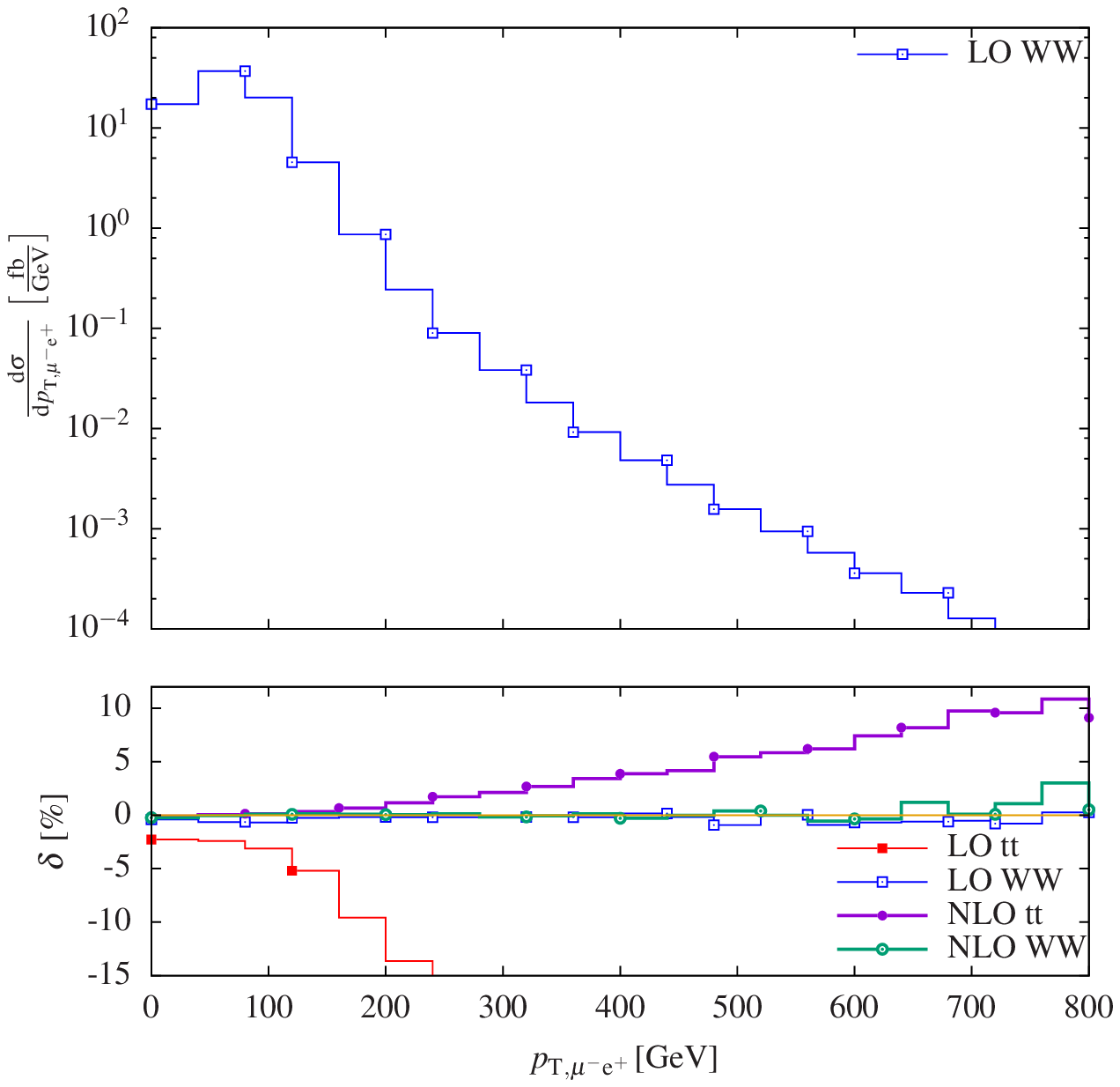}
                \includegraphics[width=0.5\textwidth]{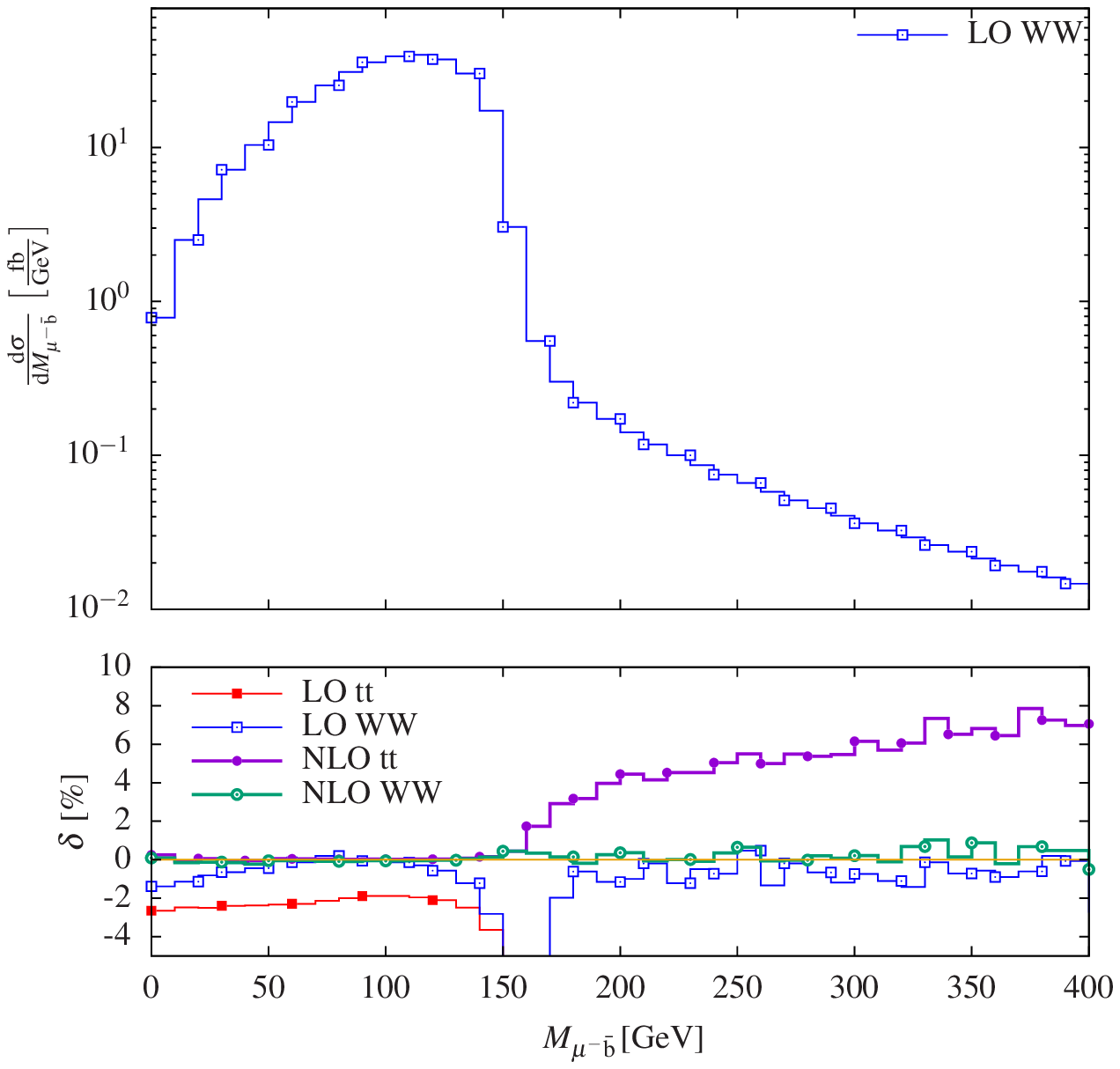}

        \caption{Differential distributions \cite{Denner:2016jyo} for ${\rm p p} \to {\rm e}^+ \nu_{\rm e} \mu^-  \bar{\nu}_\mu {\rm b} \bar{{\rm b}}$:
        transverse momentum of the muon--positron system (left) and the invariant mass of the muon and anti-bottom (right).
        Predictions for two double-pole approximations (two top quarks and two W bosons) are compared at both LO and NLO against the full calculation.}
\label{fig:DPA}
\end{figure}

\begin{figure}
\begin{center}
               \includegraphics[width=0.33\textwidth]{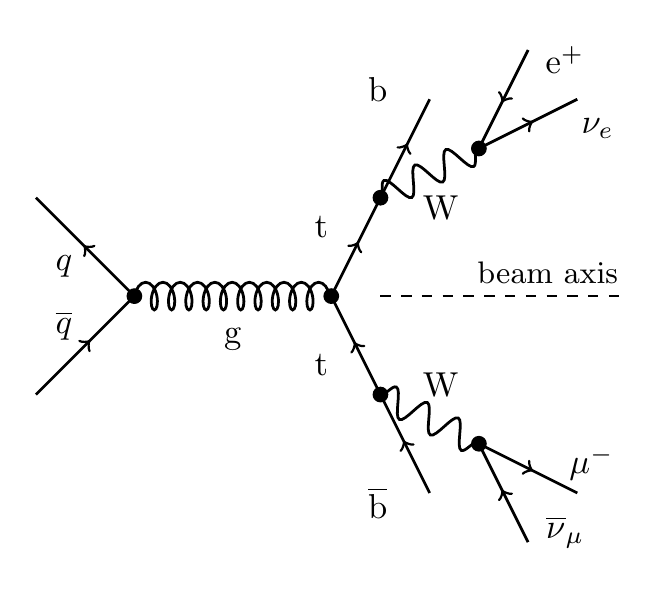}
                \includegraphics[width=0.24\textwidth]{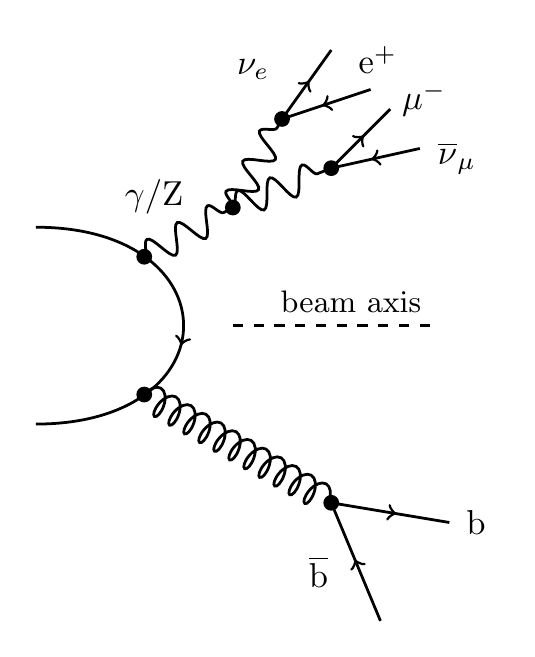}
                \includegraphics[width=0.33\textwidth]{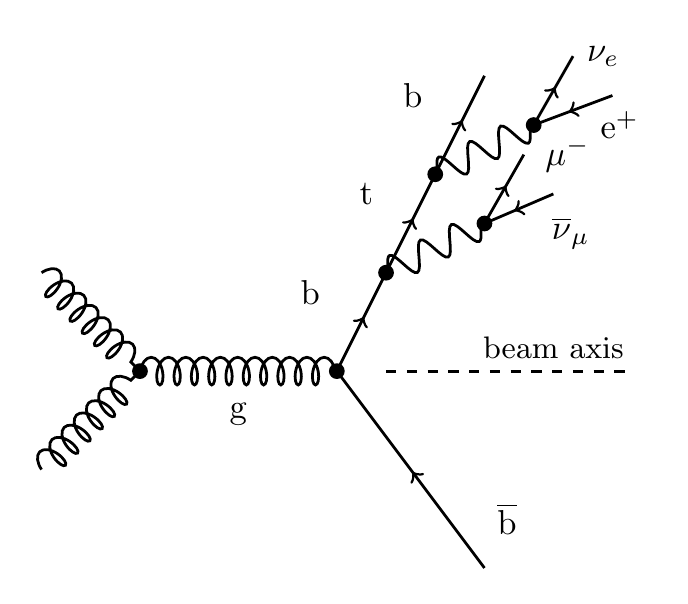}
\end{center}

\caption{Sample Feynman diagrams with two resonant top quarks (left) and no double-resonant top quarks (middle and right).
	These non-doubly resonant diagrams contribute in particular at high transverse-momentum of the muon--positron system.}
                \label{fig:diag}
\end{figure}

In order to verify the full computation, we have performed double-pole approximations (DPA) where only the doubly-resonant contributions are retained in the virtual corrections.
For this process we have devised two of these approximations: one where two top quarks are required and one where two W bosons are required.
Besides providing a good check of the computation, it also allows to infer the size of the non-resonant contributions.
In Fig.~\ref{fig:DPA}, one can see the effect of neglecting non-resonant contributions at both LO and NLO (at NLO the approximation is only applied to the virtual corrections).
In the transverse-momentum distribution of the muon--positron system, it is clear that requiring two top quarks is a particularly bad approximation for large lepton-pair transverse momenta.
At low transverse momentum, the process is dominated by back-to-back top pair production as depicted on the left of Fig.~\ref{fig:diag}, and the \emph{tt} DPA is reasonable.
For high transverse momentum however, non-resonant contributions such as the ones in the middle and right of Fig.~\ref{fig:diag} become relevant.
Such contributions are neglected in the \emph{tt} DPA explaining why it is failing to describe appropriately the high transverse-momentum region.
Hence, non-resonant contributions can be sizeable in the tail of distributions at the LHC.
In Fig.~\ref{fig:DPA}, the invariant mass of the muon and anti-bottom system is shown.
This distribution is particularly sensitive to the top-mass value \cite{Denner:2010jp,Heinrich:2017bqp,Bevilacqua:2017ipv}.
In particular, at the threshold $M_{\mu^- \bar {\rm b}}^2 = M^2_{\rm t} - M^2_{\rm W} \simeq \left(154 \; {\rm GeV} \right)^2$, none of the approximations is describing the full result properly.
This demonstrates that for arbitrary distributions (especially the ones displaying kinematic boundaries), one should only rely on the full calculation.

\section{Off-shell production of top-antitop pairs in association with a Higgs boson}

\begin{figure}
                \includegraphics[width=0.5\textwidth]{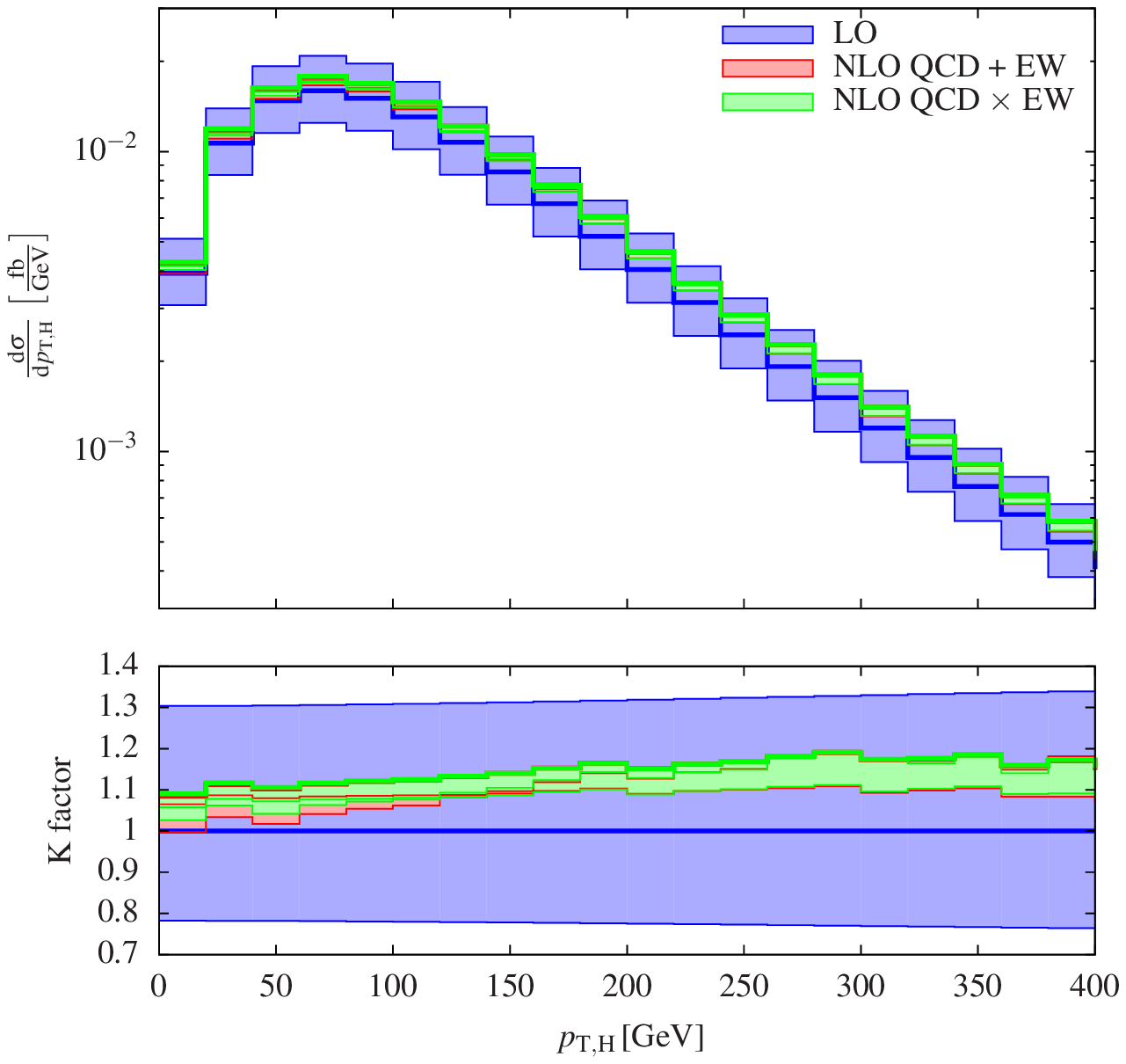}
                \includegraphics[width=0.5\textwidth]{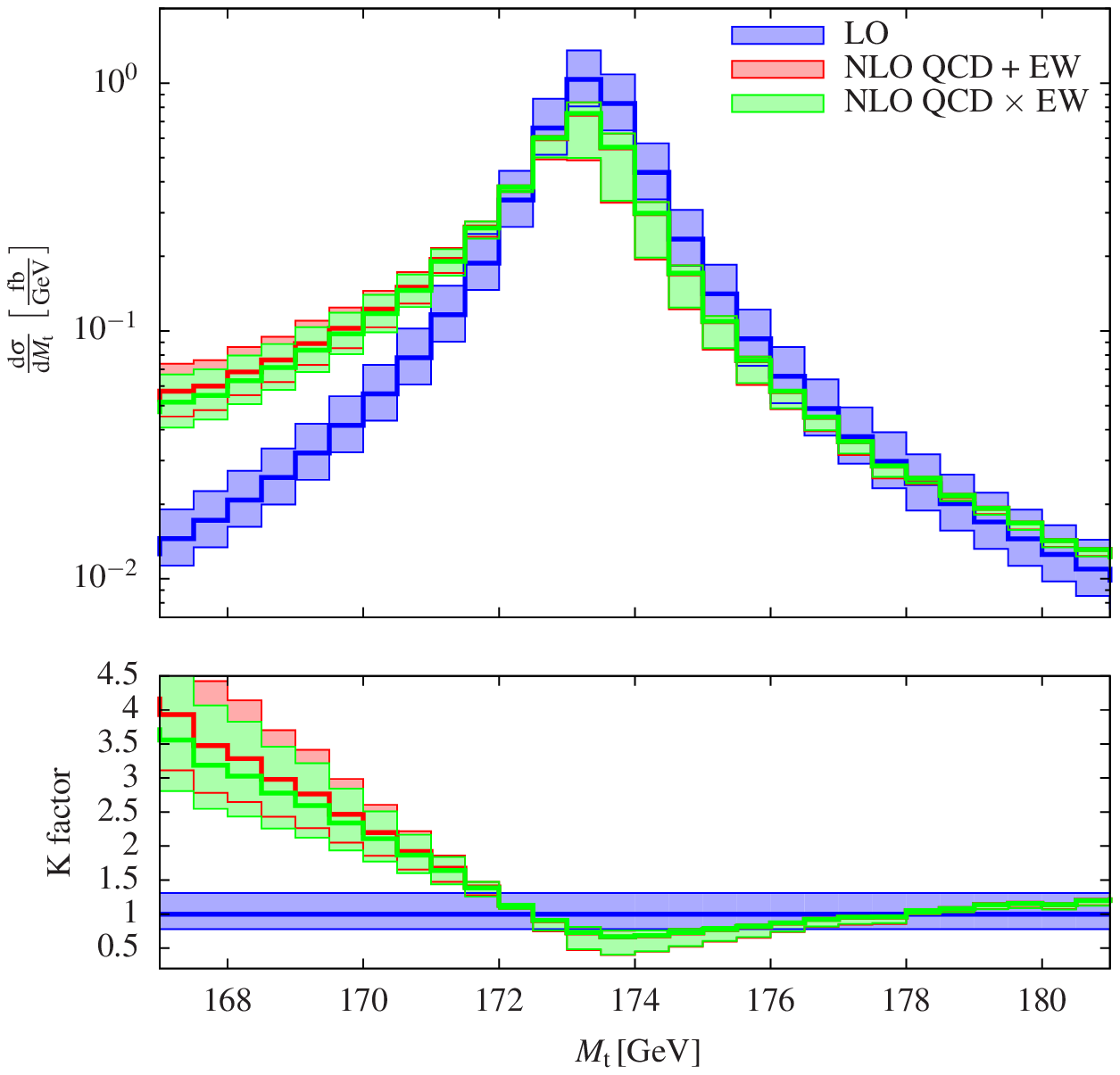}
 
        \vspace*{-3ex}
        \caption{Differential distributions \cite{Denner:2016wet} for ${\rm p p} \to {\rm e}^+ \nu_{\rm e} \mu^-  \bar{\nu}_\mu {\rm b} \bar{{\rm b}} {\rm H}$:
        transverse momentum of the Higgs boson (left) and the invariant mass of the top quark (right).
        The additive and multiplicative combination of the NLO QCD and NLO EW corrections are shown.}
                \label{fig:comb}
\end{figure}

We turn to the NLO corrections to the process ${\rm p p} \to {\rm e}^+ \nu_{\rm e} \mu^-  \bar{\nu}_\mu {\rm b} \bar{{\rm b}} {\rm H}$, considering LO contributions of order $\alpha^2_{\rm s} \alpha^{5}$ and NLO QCD and EW corrections of order $\alpha^3_{\rm s} \alpha^{5}$ and $\alpha^2_{\rm s} \alpha^{6}$, respectively.
Instead of showing the NLO EW corrections \cite{Denner:2016wet} that are comparable to those of the previous process, we present results for the combination of these NLO EW corrections with the NLO QCD ones.
The latter have already been computed in Ref.~\cite{Denner:2015yca}, and we have re-computed them in order to combine them in a common set-up.
For the combination of the NLO corrections, we have followed two prescriptions, namely the additive and the multiplicative prescription,
\begin{equation}
 \sigma^{\mathrm{NLO}}_{\mathrm{QCD+EW}} = \sigma^{\mathrm{Born}} + \delta \sigma^{\mathrm{NLO}}_{\mathrm{QCD}} + \delta \sigma^{\mathrm{NLO}}_{\mathrm{EW}} 
\end{equation}
and
\begin{equation}
 \sigma^{\mathrm{NLO}}_{\mathrm{QCD}\times\mathrm{EW}} = \sigma^{\mathrm{NLO}}_{\mathrm{QCD}} \left( 1 + \frac{\delta \sigma^{\mathrm{NLO}}_{\mathrm{EW}}}{\sigma^{\mathrm{Born}}} \right)
 = \sigma^{\mathrm{NLO}}_{\mathrm{EW}} \left( 1 + \frac{\delta \sigma^{\mathrm{NLO}}_{\mathrm{QCD}}}{\sigma^{\mathrm{Born}}} \right) ,
\end{equation}
respectively.
The quantities $\delta \sigma$ denote the respective absolute corrections.
The numerical results for the two combinations are shown at the level of differential distributions in Fig.~\ref{fig:comb}.
In the transverse-momentum distribution of the Higgs boson, the two combinations are hardly distinguishable as this distribution is dominated by the QCD corrections.
On the other hand, for the invariant mass of the top quark where EW corrections are large below the resonance, the two combinations differ substantially.
This difference can be understood as an estimate of the missing mixed higher-order corrections.

\section{Conclusion}

In this proceedings contribution we have reviewed few recent results on NLO corrections for processes involving off-shell top quarks.
The first of these processes is the production of two top quarks that decay leptonically, \emph{i.e.}\ ${\rm p p} \to {\rm e}^+ \nu_{\rm e} \mu^-  \bar{\nu}_\mu {\rm b} \bar{{\rm b}}$, where NLO EW corrections have been presented \cite{Denner:2016jyo}.
In particular, results emphasising the impact of non-resonant contributions have been shown.
The second process is the very same process with an additional Higgs boson, \emph{i.e.}\ ${\rm p p} \to {\rm e}^+ \nu_{\rm e} \mu^-  \bar{\nu}_\mu {\rm b} \bar{{\rm b}} {\rm H}$.
For this process NLO EW \cite{Denner:2016wet} corrections have been combined with pre-existing NLO QCD corrections \cite{Denner:2015yca} in order to provide state-of-the art predictions.

\acknowledgments 

We are grateful to Robert Feger for supporting {\sc MoCaNLO}.
The work of A.D. and M.P. was supported by the Bundesministerium f\"ur Bildung und Forschung
(BMBF) under contract no. 05H15WWCA1 and the work of J.-N.L. by the Studienstiftung
des Deutschen Volkes. The work of S.U. was supported in part by the European Commission
through the ``HiggsTools'' Initial Training Network PITN-GA-2012-316704.

\end{document}